\documentclass[useAMS,usenatbib]{mn2e}
\usepackage{epsf, epsfig}

\def\be{\begin{equation}}
\def\ee{\end{equation}}

\title[Quasar candidate selection and photometric redshift estimation]
{Quasar candidate selection and photometric redshift estimation based on SDSS 
and UKIDSS data}
\author[]
{Xue-Bing Wu\thanks{E-mail: wuxb@bac.pku.edu.cn} and
Zhendong Jia\\
Department of Astronomy, Peking University, Beijing 100871, P. R. China
}

\begin{document}

\date{Accepted 2010 April 07. Received 2010 March 02; in original form 2009 November 18}

\pagerange{\pageref{firstpage}--\pageref{lastpage}} \pubyear{2010}

\maketitle

\label{firstpage}

\begin{abstract}
We present a sample of 8498 quasars with both SDSS $ugriz$ optical and UKIDSS 
$YJHK$ near-IR photometric data. With this sample, we obtain the median 
colour-z relations based on 7400 quasars with magnitude uncertainties less than 
0.1mag in all bands. By analyzing the quasar colours, we propose an empirical 
criterion in the $Y-K$ vs. $g-z$ colour-colour diagram to separate stars and 
quasars with redshift $z<4$, and two other criteria for selecting high redshift
quasars. Using the SDSS-UKIDSS colour-z relations, we estimate
the photometric redshifts of 8498 SDSS-UKIDSS quasars, and find that 85.0\% of 
them are consistent with the spectroscopic redshifts within $|\Delta z|<0.2$, 
which leads to a significant increase of the photometric redshift 
accuracy than that based on the SDSS colour-z relations only. As two tests, we compare 
our colour selection criterion with a small UKIDSS/EDR quasar/star sample and a sample
of 4671 variable sources in the SDSS Stripe 82 region with both SDSS and UKIDSS 
data. We find that they can be clearly divided into two classes (quasars 
and stars) by our criterion in the $Y-K$ vs. $g-z$ plot. We select
3834 quasar candidates from the variable sources with 
$g<20.5$ in Stripe 82, 826 of them 
being SDSS quasars and the rest without SDSS spectroscopy. We 
estimate the photometric redshifts for 3519 quasar candidates with all UKIDSS
 $YJHK$ data and find an accuracy of 87.5\% within $|\Delta z|<0.2$ with 
the spectroscopic redshifts of 819 SDSS-UKIDSS identified quasars among them. 
We demonstrate that even at the same spectroscopy limit as SDSS, 
with our criterion we can at least partially recover the missing quasars with
$z\sim2.7$ in SDSS. The SDSS identified quasars only take a small fraction
(21.5\%) of our quasar candidates selected from the variable sources  
in Stripe 82, indicating that a deeper spectroscopy is very promising
in producing a much larger sample of quasars than SDSS. The implications of 
our current results to the future Chinese LAMOST quasar survey are also 
discussed.

\end{abstract}

\begin{keywords}
catalogues - surveys - quasars:emission lines - quasars:general
\end{keywords}

\section{Introduction}

The number of quasars has increased substantially in the last decade
primarily due to two large optical sky surveys, namely the Two Degree
Field (2DF) survey (Boyle et al. 2000) and the Sloan Digital Sky Survey 
(SDSS)(York et al. 2000). 2DF has 
obtained spectroscopy of more than 20,000 quasars (Croom et al. 2004), 
and SDSS has identified more
than 100,000 quasars (Schneider et al. 2007; Abazajian et al. 2009). 
Quasar candidates in these surveys were mainly
selected based on their optical colours derived from the photometric data. 
2DF mainly selected lower redshift ($z<2.2$) quasars with UV-excess 
($u-b_J<$-0.36)(Smith et al. 2005), while SDSS adopted
a multi-band optical colour selection method for quasar selection mainly by
excluding the point sources within the stellar locus of the colour-colour 
diagrams (Richards et al. 
2002). 90\% of SDSS quasars have lower redshift ($z<2.3$), although some 
dedicated methods were also developed for discovering high redshift quasars 
(Fan et al. 2001a,b; Richards et al. 2002). The lower efficiency
in identifying quasars with redshift between 2 and 3 is obvious in SDSS
(Schneider et al. 2007), because these quasars usually have similar optical 
colours as stars and thus are mostly excluded by the SDSS quasar candidate 
selection algorithm (Warren et al. 2000). 

In order to obtain a more complete sample of quasars, obviously we need to 
improve the previous quasar candidate selection methods. An important way to 
do this has been suggested by using the infrared K-band excess to identify
the `missing' quasars with redshift around 2.7 based on the UKIRT
(UK Infrared Telescope) Infrared Deep Sky Survey (UKIDSS) (Warren et al. 2000;
Hewett et al. 2006; Maddox et al. 2008).\footnote{The UKIDSS project is defined in 
Lawrence et al. (2007). UKIDSS uses the UKIRT Wide Field Camera (WFCAM; 
Casali et al. 2007) and a photometric system described in Hewett et al. 2006. 
The pipeline processing and science archive are described in 
Hambly et al. (2008).} Although the $z\sim$2.7 quasars have
similar optical colours as stars, they are more luminous in the infrared
K-band. In addition, combining the optical colours in SDSS with the infrared 
colours in UKIDSS, it should be more efficient to separate stars from  both
lower redshift quasars ($z<$3)(Chiu et al. 2007) and higher redshift ones 
($z>$6)
(Hewett et al. 2006) in the colour-colour diagrams. This advantage needs
to be confirmed by future spectroscopic observations on the larger sample of
quasar candidates selected with the SDSS and UKIDSS photometric colours.

Multi-band colours are also crucial for the photometric redshift estimations
of quasars. Based on the SDSS colour-z relations, about 70\% of the 
photometric redshifts of quasars are consistent with the 
spectroscopic ones with the difference $|\Delta z|<0.2$ (Richards et al. 2001;
Wu et al. 2004; Weinstein et al. 2004). This percentage can increase to 85\% or higher
 if the extra GALEX UV photometric data (Ball et al. 2007) or Spitzer IRAC data 
(Richards et al. 2009b) are added. It has been demonstrated that
the photo-z accuracy should be improved if combining the UKIDSS infrared 
photometry with the SDSS optical photometry for galaxies (Maddox et al. 2008).
Such improvement can be also expected for quasar candidates with both
SDSS and UKIDSS photometric data, but has not been confirmed by the photometric 
redshift estimations on a large SDSS-UKIDSS quasar sample so far. 
In addition, the reliable 
photometric redshift estimations based on multi-band colours are also important for 
preparing the quasar candidates for the future spectroscopic surveys (Richards 
et al. 2009a).   
 
Large quasar survey is one of the key projects of the Chinese Large Sky Area 
Multi-Object Fibre Spectroscopic Telescope (LAMOST), which is a novel 
reflecting
Schmidt telescope with 4 meter effective mirror size, 20 square degree field
of view (FOV) and 4000 fibres (Su et al. 1998). As the most efficient 
optical spectroscopic telescope
in the world, LAMOST finished its main construction in 2008
and has entered the commissioning phase. A pilot survey and the regular survey
have been planned in 2010 and 2011-2015 respectively. Unlike SDSS, LAMOST does
not have its own photometric survey. Therefore, the input catalogue of
LAMOST quasar survey will largely
rely on the photometric data from other existing surveys. Because of the large
overlap of the surveyed area between SDSS and UKIDSS, a combination
of SDSS and UKIDSS photometric data is expected to help us to efficiently
select a large catalogue of quasar candidates and provide reliable
photometric redshifts for the LAMOST quasar survey.   

In this study, we first present a large quasar sample with both SDSS 
and UKIDSS data, and then investigate whether there is a more efficient 
selection criterion for quasars and
whether the combination of UKIDSS data can improve the photometric redshift 
estimations. As two tests, the results are applied to a small
UKIDSS/EDR quasar/star sample and a sample of the variable
 sources in the SDSS stripe 82 region. The implications of our study to the
future LAMOST quasar survey are also discussed.         

\begin{table*}
\caption{Parameters of 8498 SDSS-UKIDSS quasars}
\centering
\begin{scriptsize}
\begin{tabular}{ccccccccccccc}\\ \hline
SDSS & SDSS & Redshift & SDSS-UKIDSS & $u$ & $g$ & $r$ & $i$ & $z$ & $Y$ & $J$ & $H$& $K$\\ 
RA & Dec & &Offset($''$)&&&&&&&&&\\ 
\hline
   0.0498394 &  0.0403587 &     0.479 &     0.1452 &       17.10 &       17.88  &   17.70     &     17.41  &   17.20   &  16.97 &  16.70 &  15.96  & 15.03 \\
   0.1625516&  -0.3010664 &      2.125  &    0.2016  &     18.16   &    19.04 &   18.64   &   18.44  &       18.21 &    18.07 &    17.76  &  17.43  &    16.75  \\  
   0.2212764&  -0.6201625 &      1.321  &    0.3712   &     17.80  &    18.59  &    18.31   &    18.10  &     18.03  &    17.94 &   17.82  &   17.41 &   16.81  \\
   0.2371434 & -1.0693223  &     2.106  &    0.2559  &     19.36 &     20.09  &     19.36   &     18.80  &     18.24 &     18.28 &    18.02  &      17.35   &    16.52  \\   
   0.2426759 & -0.7795238 &      1.897  &    0.6057 &      17.75 &     18.82  &     18.64   &    18.12   &      17.85 &   17.72  &    17.60  &   17.19 &  16.58 \\    
\hline
\end{tabular}\\
Note: All magnitudes are in Vega system. The full table of 8498 quasars is available in the electronic version of the paper.
\end{scriptsize}
\end{table*}

\section{A sample of quasars with both SDSS and UKIDSS data}
We cross-identify all quasars in SDSS Data Release 7 (DR7) with the UKIDSS Data
Release 3 (DR3), by finding the closest counterparts within 3$''$ between the 
positions in two surveys and requesting all detections  in both SDSS 
ugriz and UKIDSS YJHK bands for each quasar. To do the cross-identifications, 
we use the CrossID form available at the UKIDSS WFCAM Science Archive 
website\footnote{http://surveys.roe.ac.uk:8080/wsa/crossID\_form.jsp/},
and use only the data in UKIDSS Large Area Survey (LAS) in order to avoid the 
mis-identifications in the crowded fields with lower
Galactic latitudes. This results in a sample of 8498 quasars
with both SDSS and UKIDSS data.     

\begin{figure}
\epsfig{figure=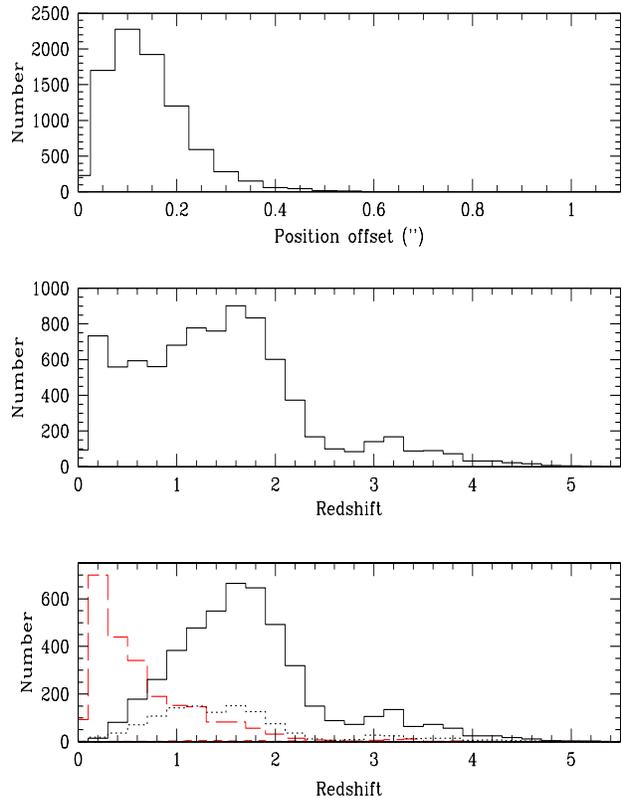, width=8.6cm, height=11.cm, angle=0}
\caption{Upper panel: Histogram of the position offsets between SDSS and UKIDSS for 
8498 quasars. Middle panel: Histogram of the quasar redshifts. Lower panel: Histograms
of the redshifts  of quasars with different UKIDSS mergedclass parameter. Solid, 
dotted, long dashed and short dashed histograms correspond to mergedclass=-1 (star), -2 (probable 
star), +1 (galaxy) and -3 (probable galaxy), respectively.}
\end{figure}

In the upper panel of Fig.1, we show the distribution of the position offsets 
between SDSS and UKIDSS for 8498 quasars. Although we use the search radius 
3$''$ for identifying the UKIDSS counterparts of SDSS quasars, we find  
that 99.6\% of the closest counterparts are within 0.5$''$ of the SDSS 
positions. Only 33 identifications
have offsets larger than 0.5$''$, and only three of them have the offset 
larger than 1$''$. Therefore, considering the subarcsecond position accuracy
 in both SDSS and UKIDSS photometry, the chance probability of
misidentification should be very small. The middle panel of Fig.1 shows
the redshift distribution of 8498 quasars. The redshifts are distributed
in a range from 0.01 to 5.29, with 7468 quasars (87.9\% of the total sample)
having $z<$2.3 and 1030 quasars (12.1\%) having $z>$2.3. Only five quasars have
redshifts larger than 5. This redshift distribution is quite similar
as that of the whole quasar sample in SDSS (Schneider et al. 2007).
The deficiency of $z\sim2.7$ quasars can be also clearly observed. In the 
lower panel of Fig. 1 we show the redshift distribution of the quasars with
different UKIDSS mergedclass parameter. Among the 8498 SDSS-UKIDSS quasars,
4905, 1181, 2383 and 29 quasars are classified with UKIDSS mergedclass=-1 (star), 
-2 (probable star), +1 (galaxy) and -3 (probable galaxy), respectively.
It is clear that most quasars with $z<0.5$ are classified as extended in
UKIDSS morphology, which confirms the significant contributions of host galaxy
light to the quasar emissions in near-IR (Maddox \& Hewett 2006; Chiu et al. 2007). 
 However, the fraction of quasars classified as point sources in UKIDSS 
morphology becomes dominant at $z>1$. This is 
understandable because it is hard to detect the host galaxies of quasars at high 
redshift even in the near-IR band.

In Fig.2 and Fig.3 we show the distributions of magnitudes and magnitude 
uncertainties of 8498 quasars in both
SDSS $ugriz$  bands and UKIDSS $YJHK$  bands. All magnitudes
are in Vega system throughout this paper if not specified. We convert the 
SDSS AB magnitudes to
Vega magnitudes using the following scalings (Hewett et al. 2006): 
$u=u(AB)-0.927,
g=g(AB)+0.103, r=r(AB)-0.146, i=i(AB)-0.366$ and $z=z(AB)-0.533$. The
magnitudes in SDSS are corrected for the Galactic extinctions using
the extinction map of Schlegel et al. (1998).     
The quasar magnitudes are mostly in a range from 17 to 20 in all SDSS optical 
bands and UKIDSS $Y$ and $J$ bands, from 16 to 19 in $H$ band and 
from 15 to 18 in $K$ band. 
The median magnitudes are 18.3,19.11,18.70,18.37, 18.13 for $ugriz$ and 
18.16,17.79,17.22, 16.59 for $YJHK$, respectively.
The magnitude uncertainties are typically
smaller than 0.05mag in SDSS $gri$ band, 0.1mag in $u$ band, and 0.15mag in 
SDSS $z$ and UKIDSS $YJHK$ bands. 
The median values of magnitude uncertainties are 0.03,0.01,0.01,0.01, 0.04 for
$ugriz$ and 0.03,0.03,0.04, 0.04 for $YJHK$, respectively. 
The median values of magnitudes and
magnitude uncertainties are also indicated in Fig. 2 and Fig. 3.

The parameters of this SDSS-UKIDSS sample of 8498 quasars, including the coordinate, 
position offset, redshift and all magnitudes in SDSS and UKIDSS bands, 
are summarized in Table 1 (only a portion is shown here for displaying the
content, and the whole table is available in the electronic version).

\begin{figure}
\epsfig{figure=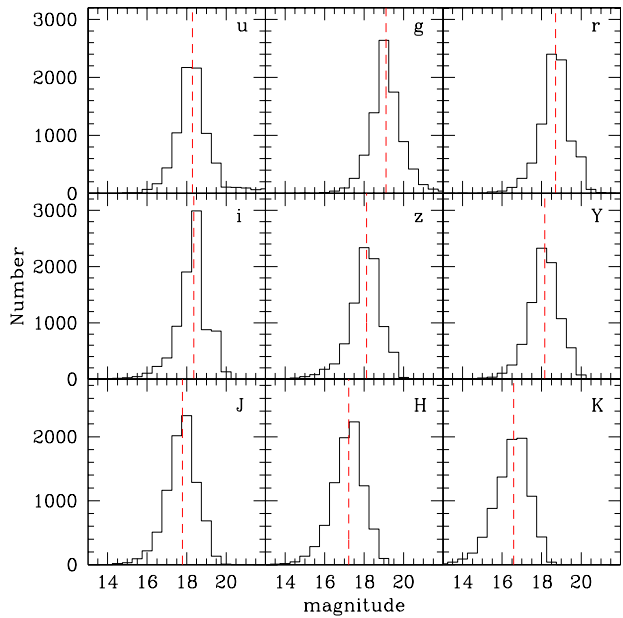, width=8.6cm, height=9cm, angle=0}
\caption{Histograms of the quasar magnitudes (in Vega system) 
in all SDSS optical bands ($ugriz$) and UKIDSS near-IR bands($YJHK$). Dashed
lines indicate the median magnitudes.}
\end{figure}

\begin{figure}
\epsfig{figure=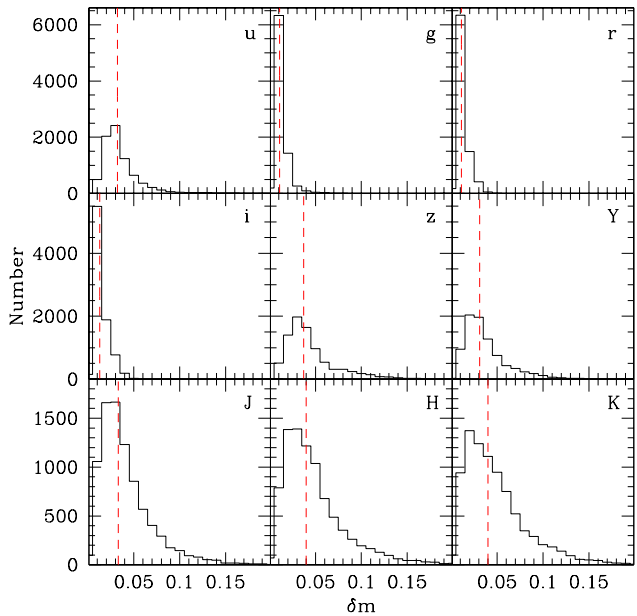, width=8.8cm, height=9cm,angle=0}
\caption{Histograms of the quasar magnitude uncertainties in all SDSS optical 
bands ($ugriz$) and UKIDSS near-IR bands ($YJHK$). Dashed lines indicate the
median values of the magnitude uncertainties.}
\end{figure}

\section {The Colour-z relations}
With the SDSS-UKIDSS sample of quasars, we can calculate the median colour-z 
relations from $u-g$ to $H-K$ colours, which have important applications in 
determining the quasar 
selection criteria and photometric redshifts. In order to obtain reliable 
colours in both optical and near-IR bands, we only use the quasars with 
magnitude uncertainties in all 9 bands ($ugrizYJHK$) less than 0.1mag. This 
reduces the quasar number from 8498 to 7400. Because the redshift range for 
our quasar sample is from 0.1 to 5.2,
we are only be able to derive the colour-z relations for $z<5.2$ quasars. We 
divide the 7400 quasars into different redshift bins with a bin size of 0.05,
0.1 and 0.2 for $z<3$, $3<z<4.5$ and $z>4.5$ respectively, and derive the 
median value of each colour at each redshift bin. We notice that the median 
$u-g$ and $g-r$ colour values derived from our sample are
quite uncertain when the redshift is larger than 3 and 4.3 respectively, due to
the small number ($<5$) of quasars in the redshift bin. Therefore, 
we only give the median colour values for $u-g$ at $z<3$ and $g-r$ at $z<4.3$. 

\begin{figure*}
\epsfig{figure=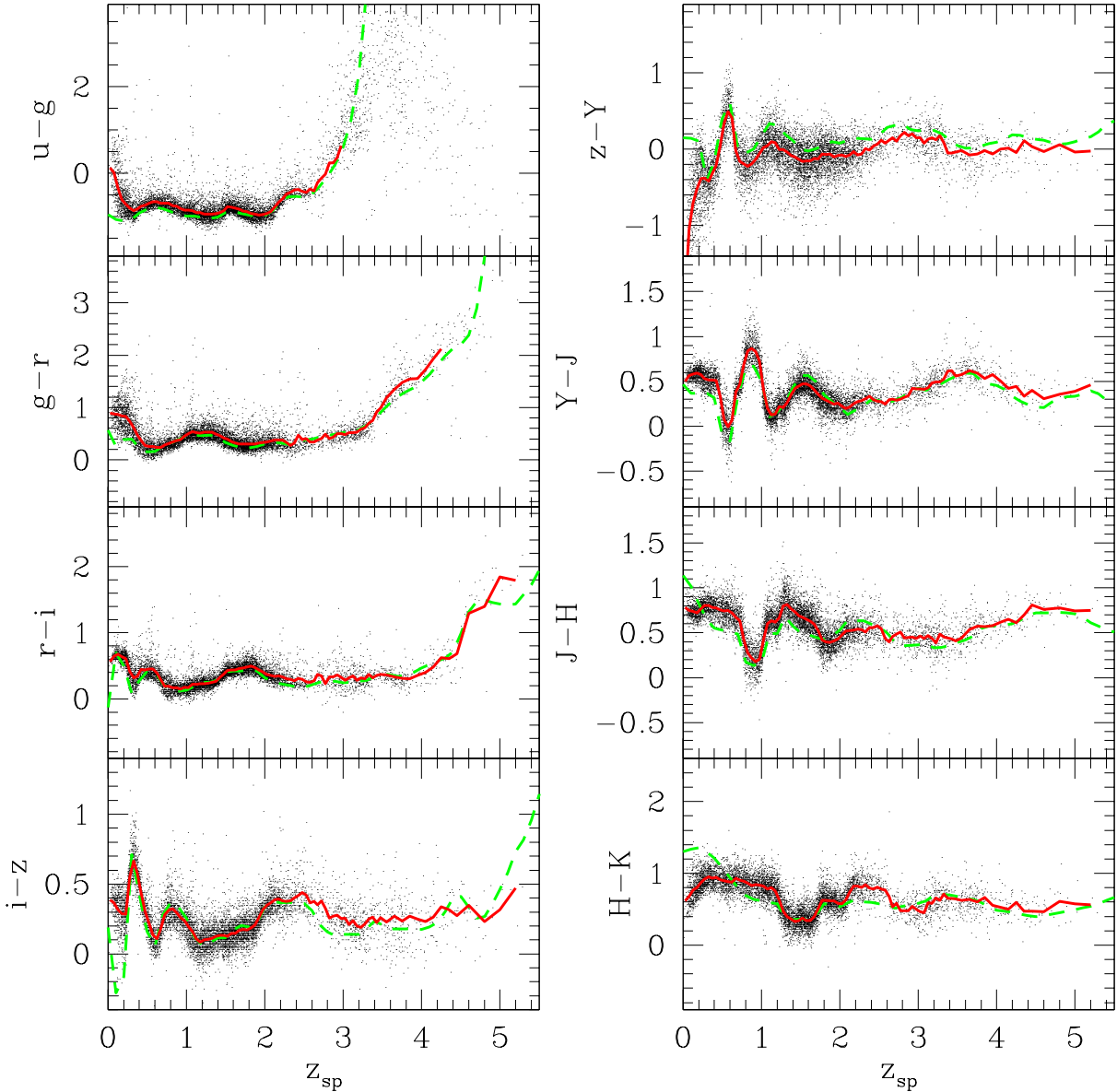, width=17cm, height=17.cm, angle=0}
\caption{The colour-z relations in optical and near-IR bands. Black dots denote 8498
SDSS-UKIDSS quasars, solid red lines represent the median colour-z relations 
derived from 7400 quasars with magnitude uncertainties less than 0.1mag. 
Dashed green lines represent the colour-z relations given by Hewett et al. 
(2006) for redshifted average model quasar spectrum.}
\end{figure*}

In a previous work, Hewett et al. (2006) presented three tables of optical/near-IR 
colours derived for redshifted blue, average and red model quasar spectra, 
which are based on the quasar model SED given in Maddox \& Hewett (2006). 
Chiu et al. (2007) plotted the colour-z relations generated from a modified 
version of the above quasar SED and compare them with the observational data 
of 2837 SDSS-UKIDSS quasars. In Fig. 4 we show the median colour-z relations 
derived from 7400 SDSS-UKIDSS quasars and the comparisons with the colours of 
all 8498 quasars in our present SDSS-UKIDSS sample.
The colour-z relations given by Hewett et al. (2006) using the mean quasar 
spectrum are also shown for comparison. The agreement between our median 
colour-z relations and those in Hewett et al. (2006) is generally very well, 
especially for all colours in optical bands and $Y-J$ colour in near-IR band. 
However, the differences
of these two colour-z relations  can be clearly seen at  $z<0.5$. Comparing 
with our median colour-z relations and the data, the colour-z relations of
Hewett et al. (2006) seem to be redder in near-IR colours (except $Y-J$) and 
bluer in optical colours for $z<0.5$ quasars. This is probably due to the continuum
shape assumption in Maddox \& Hewett et al. (2006) because the host galaxy contribution
is significant for $z<0.5$ quasars. Chiu et al. (2007) 
modified this continuum shape and obtained the colour-z relation consistent 
with observations.    
 
In Table 2 we present our median colour-z relations. For $u-g$ colour at 
$3<z<3.5$, $g-r$ colour at $4.3<z<5.0$ and all other colours at $5<z<6.4$, the 
values from the Table 26 in Hewett et al. (2006) are also added for the 
purpose of determining the quasar selection criteria and photometric redshifts 
in the following sections.

\section{Separating quasars and stars in the colour-colour diagram}
Separating quasars from stars is an important but difficult task in any
quasar survey. Usually this can be done by the optical colour selection methods
for low redshift quasars with $z<$2.3, as did by both 2DF and SDSS. 
At higher redshifts, quasars usually have similar optical colours as stars so 
the contamination from stars to the quasar candidate selection is serious.
However, as suggested by Warren et al. (2000), Chiu et al. (2007) and 
Maddox et al. (2008), this can be improved by incorporating the near-IR
colours of quasars, because quasars are relatively more luminous in K-band 
than stars even though their optical colours are almost the same in some cases
(e.g. for quasars with z$\sim$2.7).    

In some previous works, the colour-colour diagrams for quasars
and stars were analyzed by including the UKIDSS near-IR colours. Hewett et al. 
(2006) proposed to separate quasars with $6<z<7.2$ from dwarf stars in the
$i-Y$ vs. $Y-J$ diagram. Maddox et al. (2008) studied the selection criterion
of $z<4$ quasar candidates in the $g-J$ vs. $J-K$ diagram. Chiu et al. (2007)
investigated the different colour-colour diagrams in optical and near-IR bands
with a sample of 2837 SDSS-UKIDSS quasars, and found $g-r$ vs. $u-g$ diagram 
and $H-K$ vs. $J-H$ diagram are more effective in separating quasars and stars
than other colour-colour diagrams. They also proposed to use the $Y-K$ vs. $u-z$
diagram to select low redshift ($z<3$) quasars.

In section 2, we have obtained a sample of 8498 quasars, which forms
the largest sample of SDSS-UKIDSS quasars currently available. In order to
study the quasar-star separation in the colour-colour diagrams, we also 
cross-identified 8996 stars with both SDSS DR7 and UKIDSS/LAS DR3 data 
following the same procedure as for our quasar sample. These stars
are classified as `STAR' or `STAR\_LATE' in SDSS spectroscopy. All magnitudes 
of stars in SDSS $ugriz$ bands are converted to Vega magnitudes and corrected 
for the Galactic extinction.

We have produced many colour-colour diagrams with our SDSS-UKIDSS quasar and
star samples, including five colour-colour diagrams mentioned in the previous 
works mentioned above. Obviously separating quasars from stars is 
generally difficult in most colour-colour diagrams, but the colours involving 
K-band
magnitude, such as $H-K$, $J-K$ and $Y-K$, can certainly provide some helps. 
The best diagram we found to separate quasars with $z<4$ and stars is 
the $Y-K$ vs. $g-z$ diagram, which is plotted in Fig. 5. From it, we  see
that a criterion defined by $Y-K>0.46*(g-z)+0.53$, can well separate
the $z<4$ quasars and stars. This criterion is obtained by searching for
the highest efficiency in separating known SDSS-UKIDSS quasars and stars in
our samples. With this criterion, we can recover 8301 quasars in our
sample of 8498 quasars (with a percentage of 97.7\%), including 8277 of 8397 
$z<4$ quasars (with a 
completeness of 98.6\%) and 24 of 101 $z>4$ quasars (with a 
percentage of 23.8\%).
For 8996 SDSS-UKIDSS stars, only 204 of them (with a percentage of 2.3\%) 
satisfy this criterion. 
Although the percentage of mis-identifying stars as quasars is very low, we must bear in 
mind that the actual contamination rate of stars could be higher than
our estimation because SDSS only observed a tiny fraction of stars with particular 
colours spectroscopically. However, we believe that our selection criterion is still fine as
most SDSS untargetted stars (both early type and late type) should be within the stellar locus of
our $Y-K$ vs. $g-z$ diagram and would not be selected as $z<4$ quasar candidate by SDSS. 
In Fig. 5 we also show 
the predicted colours at different redshifts from the median colour-z relations 
of quasars (see section 
3), which also confirm that our criterion is applicable for identifying quasars
with $z<4$. We notice that Chiu et al. (2006) has proposed a selection 
criterion in the $Y-K$ vs. $u-z$ diagram to select quasars, but their criterion
can only be applied to $z<3$ quasars because the Ly$\alpha$ line moves out of the 
$u$-band for $z>3$ quasars. In comparison, our new criterion has the advantage
to select quasars with redshift up to 4. In addition, when we plot  
 our SDSS-UKIDSS data of quasars and
stars, we find that they are more separated in the $Y-K$ vs. $g-z$ plot than
in the $Y-K$ vs. $u-z$ plot.

Because the SDSS $z$-band photometry is not as deeper as in $ugri$ bands, our selection 
criterion in the $Y-K$ vs. $g-z$ diagram may be problematic for 
objects with $z$ magnitude fainter than 20.5 (in AB system).
Therefore, we also propose another criterion in the $Y-K$ vs. $g-Y$ diagram, 
namely,
$Y-K>0.32*(g-Y)+0.67$, to separate quasars from stars.  With this criterion,
we can recover 8315 quasars in our sample of 8498 quasars, with a percentage 
of 97.8\%, including 8276 of 8397 $z<4$ quasars (with a 
completeness of 98.6\%) and 39 of 101 $z>4$ quasars (with a 
percentage of 38.6\%), but we also select 304 of 8996 stars (with a 
percentage of 3.4\%). The star contamination is higher than  the case of
using the criterion in the $Y-K$ vs. $g-z$ diagram. In the
following study, we will mainly use the criterion in the  $Y-K$ vs. $g-z$ 
diagram for selecting $z<4$ quasars in order to keep a lower 
contamination rate of stars.  

\begin{figure}
\epsfig{figure=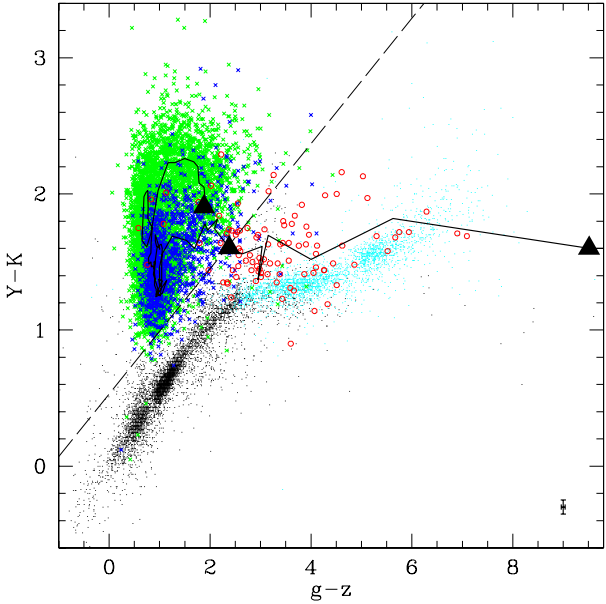, width=8.8cm, angle=0}
\caption{The $Y-K$ vs. $g-z$ colour-colour diagram of the SDSS-UKIDSS quasar and 
star samples. Green
crosses, blue crosses and red circles denote quasars with $z\le2.2$, 
$2.2<z\le4$ and $z>4$, respectively. Black dots and cyan dots denote stars 
classified as `STAR' and `STAR\_LATE' by SDSS spectroscopy. Dashed line 
indicates the separation criterion $Y-k>0.46*(g-z)+0.53$. Solid curve is 
derived from the median colour-z relation (see section 3) of quasars. 
 The solid triangles, from left to right,
 mark the colours of quasars with z=0.02, 4, and 5 respectively, derived from 
the colour-z relation. Typical error bars are
shown in the lower-right corner. }
\end{figure}

\begin{figure}
\epsfig{figure=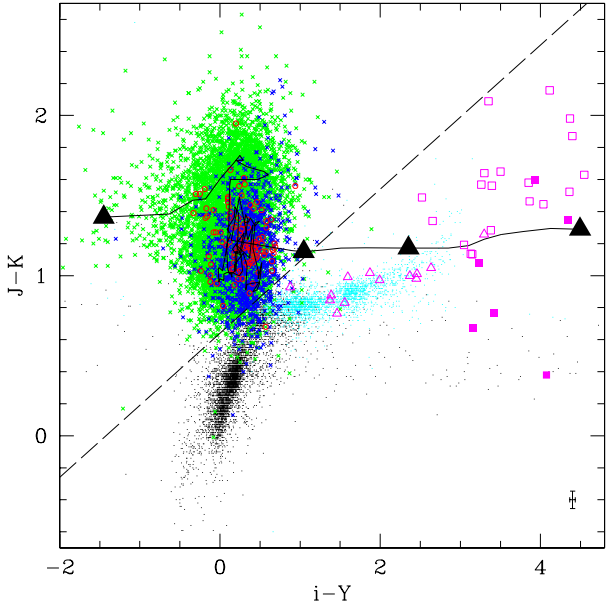, width=8.8cm, angle=0}
\caption{The $J-K$ vs. $i-Y$ colour-colour diagram of the SDSS-UKIDSS quasar and 
star samples. Open triangles, open 
squares and solid squares denote M, L and T dwarfs given in Tables 9 and 10
of Hewett et al. (2006). Other symbols have the same meaning as in Fig. 5. 
Dashed line indicates the separation criterion $J-K=0.45*(i-Y)+0.64$. Solid 
curve is derived from the colour-z relation (see section 3) of quasars. The 
solid triangles, from left to 
right, mark the colours of quasars with z=0.02, 5.3, 5.7 and 6.4 
respectively, derived from the colour-z relation. Typical error bars are
shown in the lower-right corner.}
\end{figure}

\begin{figure}
\epsfig{figure=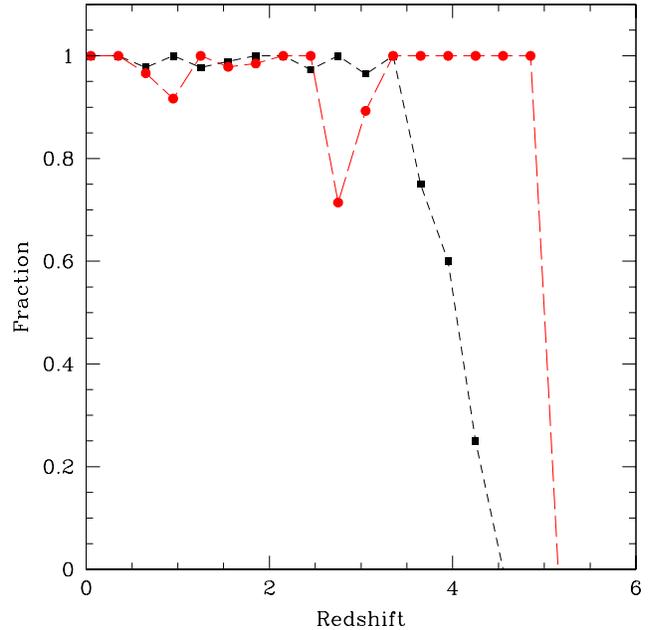, width=8.5cm, angle=0}
\caption{Fraction of the 704 FIRST radio-selected SDSS/UKIDSS quasars 
recovered by 
applying two quasar selection criteria. The short and long dashed lines 
denote the cases of using the criteria in the $Y-K/g-z$ and $J-K/i-Y$ plots, 
respectively. The redshift bin size is 0.3.}
\end{figure}

\begin{figure}
\epsfig{figure=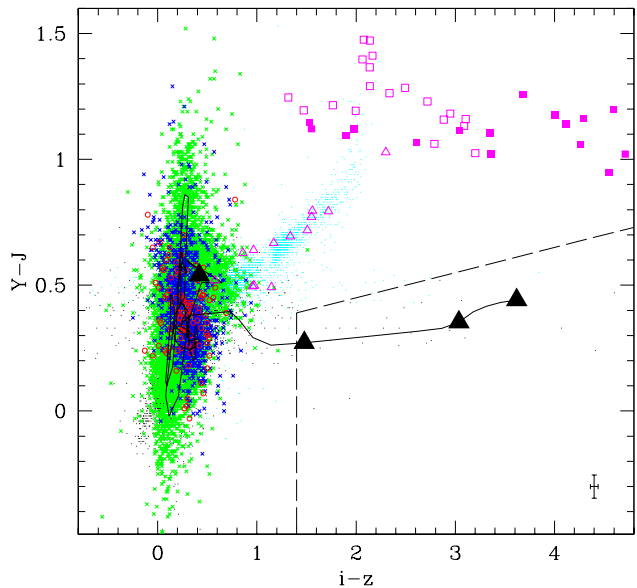, width=8.5cm, angle=0}
\caption{The $Y-J$ vs. $i-z$ colour-colour diagram of the SDSS-UKIDSS quasar and star
samples. Symbols have the same meaning as in Fig. 7.  Solid  curves are 
derived from the colour-z relation (see section 3) of quasars. The solid
triangles, from left to right, mark the colours of quasars with z=0.02, 5.6, 6.0
and 6.4 respectively, derived from the colour-z relation. Straight dashed lines 
indicate the selection criterion $i-z>1.4$ \& $Y-J<0.1*(i-z)+0.25$ for
$z>5.6$ quasars. Typical error bars are
shown in the lower-right corner.}
\end{figure}

For quasars with  $z>4$, obviously we need other selection criteria. In most
colour-colour diagrams, these high-z quasars have similar colours as late-type
stars because the strong Ly$\alpha$ line of quasars moves to the red 
part of optical spectra. By looking carefully at the various colour-colour 
diagrams, we realize that the high-z quasars and late-type stars often 
have different
$J-K$ and $i-Y$ colours. We make a $J-K$ vs. $i-Y$ colour-colour diagram (shown in
Fig.6) with the data of SDSS-UKIDSS quasars and stars. Although some
$z<4$ quasars have the same colours as stars, the $z>4$ quasars can be well
separated from stars with a criterion $J-K>0.45*(i-Y)+0.64$. In Fig. 6 we also 
show the predicted colours from the median colour-z relations of quasars 
(see section 3), which also confirm that our criterion is applicable for 
identifying quasars with $z<5.3$, although some quasars with $2.5<z<3$, whose
colours ($i-Y\sim$0.5 and $J-K\sim$0.8) are similar as stars in this colour-colour
 diagram, could be missed by 
applying this criterion. We check this criterion with our SDSS-UKIDSS sample of
quasars and stars, and find that using the criterion $J-K>0.45*(i-Y)+0.64$ 
we can recover 8289 of 8498 quasars in our sample (with a percentage of 
97.5\%), including 100 of 101 $z>4$ quasars (with a percentage of 99.0\%) 
and 8189 
of 8397 $z<4$ quasars (with a percentage of 97.5\%). Only one $z>4$ quasars
and 208 $z<4$ quasars are missed. For 8996 stars, 474 of them (with a 
percentage of 5.3\%) are mis-identified as quasars by our criterion. 
Therefore, the criterion in the $J-K$ vs. $i-Y$ diagram is better than the
criterion in the $Y-K$ vs. $g-z$ diagram for identifying high-z quasars (with 
$4<z<5.3$), although the latter is better for identifying $z<4$ quasars. 

Perhaps a better way is to combine these two criteria for identifying 
quasars with redshift up to 5.3, namely,
\be
Y-K>0.46*(g-z)+0.53, ~\rm{or}
\ee
\be
J-K>0.45*(i-Y)+0.64.
\ee
With the above criteria, we can recover 8447 of 8498 quasars in our sample 
(with a percentage of 99.4\%), including 8347 of 8397 $z<4$ quasars (with a 
percentage of 99.4\%) and 100 of 101 $4<z<5.3$ quasars (with a 
percentage of 99.0\%). However, we would also mis-identify 566 stars as 
quasars, with a percentage of 6.3\% of the total number of stars in our
current SDSS-UKIDSS sample. In this
case, we can achieve the highest efficiency in recovering both $z<4$ and $4<z<5.3$ 
quasars, but the contamination rate of stars is also higher than using
the individual criterion we proposed above.  

Because the radio quasars identified in SDSS were not selected by the 
colour-based methods (Richards et al. 2002), we can make a check with the FIRST 
radio-detected SDSS quasars to estimate the completeness of applying our 
selection criteria, similar as what has been 
done in Richards et al. (2006).  In our sample of 8498 
SDSS/UKIDSS quasars, 704 of them were detected by FIRST. With their photometric 
data in SDSS and UKIDSS, we can check how many of them can be recovered by our 
quasar selection criteria in the $Y-K/g-z$ and $J-K/i-Y$ colour-colour diagrams.
In Fig. 7 we show the fraction of recovered quasars at each redshift bin for
two criteria we proposed above for $z<4$ and $4<z<5.3$ quasars respectively. It
is clear that with our $Y-K/g-z$ criterion we can achieve the fraction higher than
95\% in each redshift bin with $z<3.5$, while with our $J-K/i-Y$ criterion we can
recover all FIRST quasars with $3.5<z<5$. Especially with our $Y-K/g-z$ criterion,
we can recover the quasars with $z\sim2.7$ very well, which seems to be difficult
in the case of using the SDSS quasar selection criteria alone (Richards et al. 
2006). Such a check supports that with our two criteria we can identify quasars 
at a completeness higher than 95\% if their redshifts are smaller than 5.

Although our current SDSS-UKIDSS quasar sample does not include any quasar with
$z>5.3$, we can explore how to improve the high-z quasar selection by using
the colour-z relations. For quasars with $z>5.3$, especially those with $z>5.7$,
from Fig. 6 we can see clearly that the contaminations of M, L and T dwarfs 
are much serious in the $J-K$ vs. $i-Y$ diagram. This is a problem already 
known for identifying high-z 
quasars (Fan et al. 2001b; Hewett et al. 2006), and the methods to solve it
all involved near-IR colours. Fan et al. (2001b) proposed a selection criterion
for $z\sim6$ quasars in the $z-J$ vs. $i-z$ diagram by requesting $i-z>2.2$ and
$z-J<1.5$. Hewett et al. (2006) mentioned that $z>6$ quasars can be identified
in the $i-Y$ vs. $Y-J$ diagram as they have bluer $Y-J$ colours than L and T 
dwarfs. Here we present another colour-colour diagram ($Y-J$ vs. $i-z$) in 
Fig. 8 and propose another selection criterion for high-z quasars: $i-z>1.4$ 
and  $Y-J<0.1*(i-z)+0.25$. It is clear that using both $i$ drop-out and near-IR 
colour can help us to separate $z>5.6$ quasars from stars and low-z quasars, as 
well as T and L dwarfs effectively, though the contaminations from some stars
are still unavoidable. Further spectroscopic observations on some high-z quasar
candidates selected with our criterion are expected.
  
\section{Photometric redshift estimation}
With the colour-z relations derived in section 3, we can estimate the 
photometric shifts of quasars by comparing the observed colours with the 
predicted colours at certain redshift by the colour-z relations. Combining SDSS 
$ugriz$ and
UKIDSS $YJHK$ magnitudes, we can obtain 8 colours from optical $u-g$ colour to 
near-IR $H-K$ colour for the quasar candidates, then using the $\chi^2$ 
minimization method we can obtain the most probable photometric redshifts for 
them. Here the  $\chi^2$ is defined as (see Wu et al. 2004):
\be
\chi^2=\sum_{ij}{\frac{[(m_{i,cz}-m_{j,cz})-(m_{i,observed}-m_{j,observed})]^2}
{\sigma_{m_{i, observed}}^2+\sigma_{m_{j,observed}}^2}},
\ee
where the sum is obtained for all 8 SDSS-UKIDSS colours, $m_{i,cz}-m_{j,cz}$ is 
the colour in the 
colour-z relations, $m_{i,observed}-m_{j,observed}$ is the observed colour 
of a quasar, $\sigma_{m_{i, observed}}$ and $\sigma_{m_{j,observed}}$ are the
uncertainties of observed magnitudes in two SDSS-UKIDSS bands.

\begin{figure}
\epsfig{figure=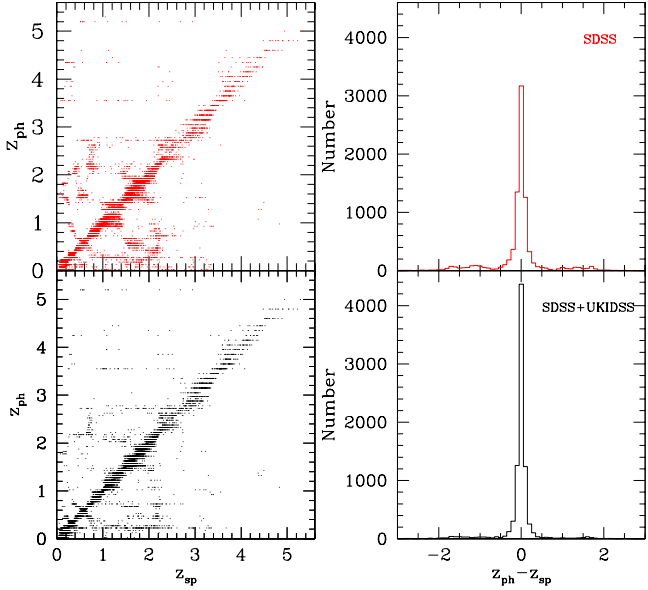, width=9cm,angle=0}
\caption{Comparison of photometric redshift estimation results of 8498 
SDSS-UKIDSS quasars for the cases of using 
SDSS colours only (upper panels) and using SDSS-UKIDSS colours (lower panels). 
$z_{ph}$ and $z_{sp}$ denote photometric and spectroscopic redshifts 
respectively.}
\end{figure}

We notice that the Ly$\alpha$ emission line moves out of the  
$u$ band and $g$ band when the redshift is larger than 3.5 and 5 respectively,  
so in our derived colour-z relation in section 3, we don't have the $u-g$ value
for $z<3.5$ and both the $u-g$ and $g-r$ values for $3.5<z<5$. This means that we
can only use 7 and 6 of 8 SDSS-UKIDSS colours if the input redshift is larger 
than 3.5 and 5 respectively. In order to compare the $\chi^2$ values for
the cases of using different number of colours at different redshifts, we
actually use the $\chi^2/N$ (where N is the number of colours and is 8,7,6
respectively for input redshift $z<3.5$,$3.5<z<5$ and $z>5$) instead of 
$\chi^2$ to determine the photometric redshift by obtaining the minimum of
$\chi^2/N$ at certain redshift. An IDL program is made for searching the
photo-z of quasars by taking the above procedures into account. 

We estimate the photometric redshifts of 8498 SDSS-UKIDSS quasars with our 
program. In order to see more clearly whether using the UKIDSS colours can 
improve the photo-z estimation, we also estimate the photometric redshifts
of these quasars using the 4 SDSS colours only. Similarly as we did above,
actually only 3 and 2 of 4 SDSS colours can be used for input redshifts larger 
than 3.5 and 5, and photometric redshift is determined by searching the minimum
of $\chi^2/N$. In Fig.9 we show the comparison of the photo-z results obtained
with SDSS and SDSS-UKIDSS colours respectively. It is obvious that using 
both SDSS and UKIDSS colours can improve the photo-z accuracy substantially.
For 8498 quasars, 7223 of them (with a percentage of 85.0\%) have photometric 
redshifts  given by the SDSS-UKIDSS colours consistent with the 
spectroscopic redshifts within $|\Delta z|\le 0.2$. 6102 of them (with a 
percentage of 71.8\%) have 
$|\Delta z|\le 0.1$. In comparison, if 
using the SDSS colours only, 6179 of quasars (with a percentage of 72.7\%) have 
photometric redshifts  consistent with the 
spectroscopic redshifts within $|\Delta z|\le 0.2$, and 4898 of them (with a 
percentage of 57.6\%) 
$|\Delta z|\le 0.1$. Therefore, by adding the 
UKIDSS $YJHK$ data, the accuracy of photo-z can be improved with a percentage
of 12.3\% for $|\Delta z|\le 0.2$ and 14.2\% for $|\Delta z|\le 0.1$, 
respectively. This substantial improvement is really what we expected. 
From Fig. 9 we notice that there are still large failures in the photo-z estimations
for some quasars, even if the UKIDSS data are added. This is mainly due to the 
photo-z degeneracy (the $\chi^2$/N minimum values are comparable at two or more 
different input redshifts) (Wu et al. 2004), and different colours of some special
quasars, such as the reddened quasars (Weinstein et al. 2004). More dedicated
approaches, such as using the additional morphology and colour information, could
help to further improve the photo-z accuracy.  
 
\section{Tests with a small UKIDSS/EDR sample and variable sources in SDSS 
Stripe 82 region}

In the two sections above, we presented the selection criteria of quasar
candidates at different redshift and the photometric redshift estimation
method. Obviously more tests on these techniques need be done with some 
existing photometric and spectroscopic samples before their applications to a 
large quasar survey.

\begin{figure}
\epsfig{figure=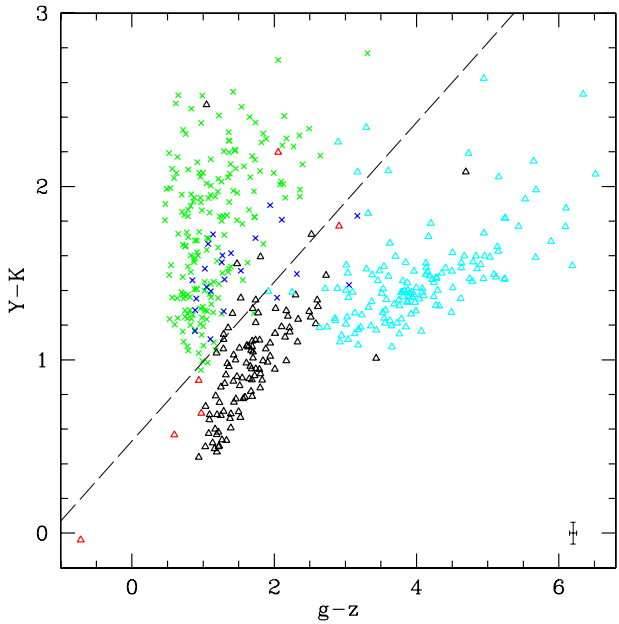, width=8.8cm,angle=0}
\caption{Testing the quasar selection criterion in $Y-K$ vs. $g-z$ colour-colour 
diagram with the
observational data in Maddox et al. (2008). Red, black, and cyan triangles
denote the A, K, and M stars respectively. Green and blue crosses denote 
quasars with $z\le2.2$ and $2.2<z\le4$. Dashed line represents our quasar 
selection criterion shown in Fig. 6. Typical error bars are
shown in the lower-right corner.}
\end{figure}

\begin{figure}
\epsfig{figure=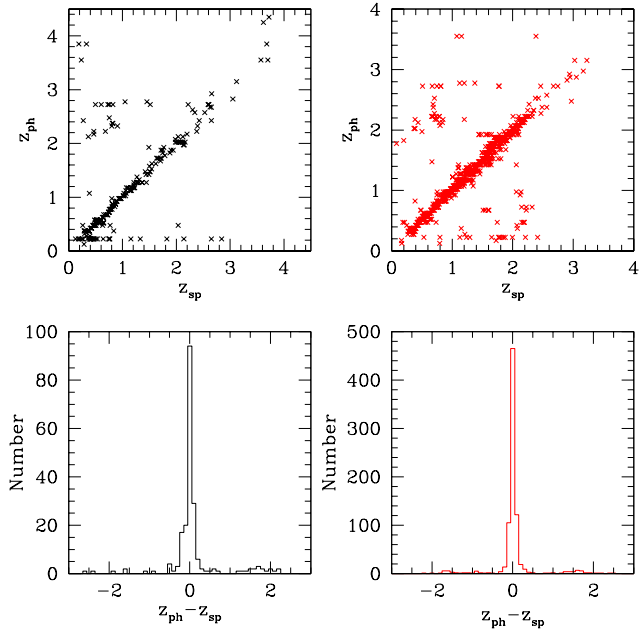, width=8.8cm,angle=0}
\caption{Comparison of the photometric redshifts with the spectroscopic 
redshifts for 209 quasars (left panels) in Maddox et al. (2008)  and 819 
SDSS identified quasars (right panels) in a sample of variable sources in
Stripe 82.}
\end{figure}

First we test our selection criteria with a sample of 209 objects classified
as QSOs and 252 objects classified as A, K, and M stars in Maddox et al. 
(2008). These objects (with K$\le17.0$) were selected from UKIDSS/LAS Early 
Data Release and observed with the AAOmega multi-object spectrograph. In Fig. 
10 we plot these objects in the $Y-K$ vs. $g-z$ diagram and compare them with
our quasar selection criterion, $Y-K>0.46*(g-z)+0.53$. Among 209 quasars,
186 of them have $z\le2.2$ and 182 can be found with our criterion, 
23 of them
have $2.2<z<4$ and 19 of them meet our criterion. Therefore, using our
proposed selection criterion we can recover 201 of 209 quasars (with a 
percentage of 96.2\%) in Maddox et al. (2008). However, 13 of 252 stars, 
including one A stars, nine K stars and three M stars, also satisfy our 
criterion. The fraction of these mis-identified
stars is 5.2\% of the total number of stars and 6.1\% of the total number of
objects satisfying our selection criterion. Because all 209 quasars in this 
sample have $z<4$, this test confirms that our criterion is indeed applicable
for identifying $z<4$ quasars. Maddox et al. (2008) indicated that 17 quasars
in their sample do not have SDSS spectra, therefore, this test also 
implies that with our criterion we are able to recover these SDSS missing 
quasars. In the left panels of Fig. 11 we show the comparison of the 
photometric redshifts estimated by our program (see section 5) with the 
spectroscopic redshifts for 209 quasars in this sample.
158 quasars have $|\Delta z|\le 0.2$ and 131 quasars have $|\Delta z|\le 0.1$,
with a percentage of 75.6\% and 63.2\% of the total 209 quasars, respectively.
This also confirms the robustness of our photometric redshift estimations.

\begin{figure}
\epsfig{figure=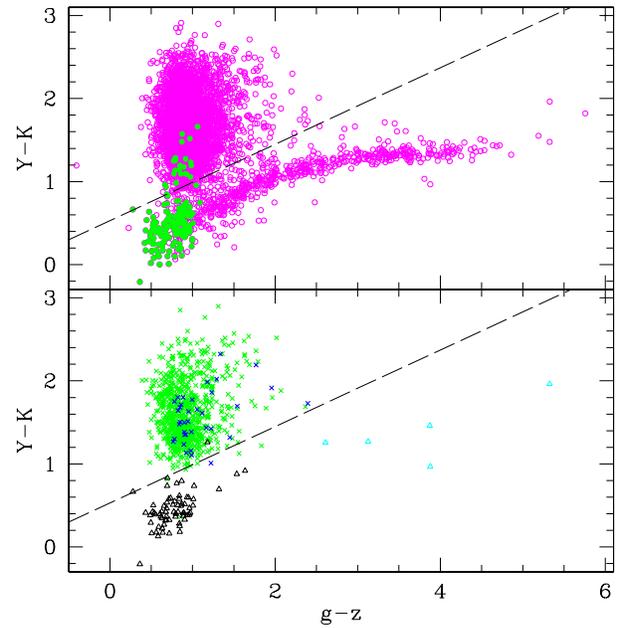, width=8.8cm,angle=0}
\caption{Upper panel: $Y-K$ vs. $g-z$ colour-colour diagram of all 4671 variable 
sources in SDSS Stripe 82 (Sesar et al. 
2007) with SDSS-UKIDSS data. Green solid circles denote the 169 RR Lyrae star 
candidates.
 Lower panel: Only the variable sources with SDSS spectroscopy are 
shown. Green and blue crosses denote the quasars with
 $z \le 2.2$ and 
$2.2<z<4$ respectively. Black and cyan triangles denote the sources classified 
as STAR and STAR\_LATE by SDSS respectively.  Dashed lines in 
both panels represent the quasar selection criterion shown in Fig. 6.}
\end{figure}

\begin{figure}
\epsfig{figure=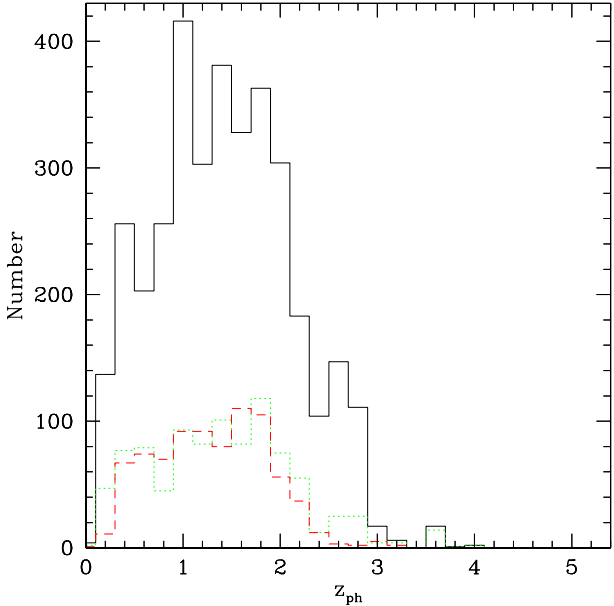, width=8.8cm,angle=0}
\caption{Histogram of photometric redshifts of 3539 quasar candidates selected
with our criterion from variable sources with $g<20.5$ in Stripe 82. Red dashed
histogram shows the spectroscopic redshift distribution of  819 SDSS identified
quasars. Green dotted histogram shows the redshift distribution of quasar 
candidates with $i_{\rm{vega}}<19.1$, which is the SDSS spectroscopic limit for
 $z<2.3$ quasars.}
\end{figure}

The second test we made is for a sample of variable sources in SDSS Stripe 82,
which is a region of 290deg$^2$ located along the celestial equator in the
southern Galactic cap and was repeatedly observed by SDSS photometry (Ivezic
et al. 2007). Sesar et al. (2007) presented 13051 candidate variable
sources with criteria $g<20.5$, $\sigma(g,r)\ge0.05$mag and $\chi^2(g,r)\ge3$
(here $\sigma(g,r)$ and $\chi^2(g,r)$ are the variability amplitude and
$\chi^2$ per degree of freedom in $g$ and $r$ bands respectively) from 
1.4 million unresolved sources in Stripe 82 observed at least 4 times
in each of SDSS $ugri$ bands. 
Following the same procedure as in section 2,
we cross-identify their near-IR counterparts with the UKIDSS/LAS DR3 and find
4671 of 13051 sources with detections in both $Y$ and $K$ bands. In the upper 
panel of Fig. 12 we plot
the colours of these 4671 variable sources in the $Y-K$ vs. $g-z$ diagram and 
compare them with our quasar selection criterion. 
The SDSS magnitudes of these 4671 variable sources are taken to be the mean 
magnitudes given in Sesar et al. (2007). It can be clearly seen that
these variable sources are divided into two categories approximately by
our criterion. 3834 of these 4671 sources (with a percentage of 82.1\%) 
satisfy our criterion, while 837 sources, which do not satisfy our criterion,
obviously locate in the stellar locus as we see in Fig. 5 and Fig. 10. 
In Fig. 12 we also plot 169 sources in this sample, which meet the selection 
criteria for RR Lyrae stars (Ivezic et al. 2005; Sesar et al. 2007). 145 
sources of these 169 RR Lyrae candidates (with a percentage of 85.8\%) locate 
in the stellar locus, and we believe that they are most likely real RR Lyrae 
stars. However, 24 of these RR Lyrae candidates  meet our quasar selection 
criterion. Future spectroscopic observations on these 24 
RR Lyrae candidates will help us to clarify their nature.

Among 4671 variable sources with both SDSS and UKIDSS $Y$ and $K$ band data, 
834 of them have been identified by SDSS spectroscopy as quasars with redshifts
$0.1<z<3.3$. In the lower panel of Fig. 12 we also compare their colours with
our selection criterion. We can clearly see that 826 of them (with a percentage
of 99.0\%) satisfy our criterion, and only 8 quasars are exceptions (5 of them 
actually locate very close to our criterion line in Fig. 12). 72 of 4671 
sources have been identified as `STAR' and `STAR\_LATE' by SDSS spectroscopy 
respectively.
Two of them (with a percentage of 2.8\%) satisfy our quasar selection criterion,
and the rest 70 stars clearly locate in the stellar locus in Fig. 12. This
again proves the robustness of our quasar selection criterion. Among 834 SDSS
identified quasars, 819 of them have UKIDSS data in all $YJHK$ bands. We 
estimate their photometric redshifts with our photo-z program and compare their
photometric redshifts with spectroscopic ones. The results are shown in the
right panels of Fig. 11. 717 quasars have $|\Delta z|\le 0.2$ and 627 quasars 
have $|\Delta z|\le 0.1$,
with a percentage of 87.5\% and 76.6\% of the total 819 quasars, respectively.
This tight agreement demonstrates again that using both SDSS and UKIDSS colours 
we can obtain more accurate photometric redshifts for quasars.

Among 3834 quasar candidates selected with  our selection
criterion in the $Y-K$ vs. $g-z$ diagram from 4671 variable sources in Stripe 82,
3539 sources have all UKIDSS $YJHK$ detections. We estimate their photometric 
redshifts with our photo-z program and the results are shown in Fig. 13. 3519 
quasar candidates have $z_{ph}\le 3.5$, 20 candidates have $3.5<z_{ph}<5$, and 
none candidate has $z_{ph}>5$. This is understandable because our selection
criterion in the $Y-K$ vs. $g-z$ diagram is mainly for selecting $z<4$ quasars
(we also used our second criterion in the $J-K$ vs. $i-Y$  diagram to select 
possible $z>4$ quasar candidates, but this only contributed 61 additional 
candidates). 
In Fig. 13 we also show the spectroscopic redshift distribution of 819 SDSS
identified quasars with all UKIDSS $YJHK$ detections, and the photometric
redshift distribution of our quasar candidates with $i$ magnitude in Vega 
system $i_{\rm{AB}}<19.1$, which is the SDSS spectroscopic limit for 
$z<2.3$ quasars. Comparing these two redshift distributions, we
realize that even at the same magnitude limit, with our criterion we can probably
select more quasars with $z\ge2$ than SDSS. Especially for $z\sim2.7$ quasars,
which are largely missing in SDSS because of their similar colours as stars, can
be at least partially recovered with our quasar selection criterion.   
For our 3834 quasar candidates (with $Y$ and $K$ detections), 
only 826 of them were identified
as quasars by SDSS spectroscopy, which is only a small fraction (21.5\%) of
our quasar candidates selected from variable sources in Stripe 82. This is
mainly because the SDSS spectroscopic limit is $i_{\rm{AB}}<19.1$ for low-z
quasars, which
is shallow than $g_{\rm{AB}}<20.5$ for selecting variable sources in Stripe 82
(Ivezic et al. 2007; Sesar et al. 2007). We also notice that 207 sources 
of our quasar candidates in Stripe 82 have been recently identified as 199 new $z<3$ 
quasars, 
2 white dwarfs and 6 unclassified objects by the 2dF-SDSS LRG and QSO survey 
(2SLAQ) (Croom et al. 2009).
Therefore, deeper spectroscopy than SDSS, such as LAMOST quasar survey 
which is expected to reach a limit of $i_{\rm{AB}}=20.5$, will hopefully 
produce a much larger sample of quasars in the future. 

In addition, we found
that 837 of 4671 variable sources with both SDSS and UKIDSS ($Y$ and $K$) data
do not satisfy our quasar selection criterion in the $Y-K$ vs. $g-z$ diagram. 
They clearly locate in the stellar locus in this diagram and most of them are
probably stars. These 837 star-like sources only take a small fraction (17.9\%) 
of our 4671 SDSS-UKIDSS variable sources in Stripe 82, which confirms that 
the variability study is very helpful for discovering quasars (Sesar et al. 2007).       
Our results suggest that with the near-IR photometric data we can further improve
the accuracy of selecting quasars from optical variability study. This is
probably important for the variability study in the future large photometric surveys, 
such as that with the Pan-STARRS 
(Kaiser et al. 2002) and Large Synoptic Survey Telescope (Tyson 2002).  

\section{Discussions}
Quasar candidate selections largely rely on the multi-band photometric
colours. We have demonstrated that combining the UKIDSS $YJHK$ data with SDSS
$ugriz$ data will lead to significant improvements in both quasar selection
efficiency and photometric redshift accuracy. With our quasar selection
criteria involving both SDSS and UKIDSS colours, we can probably select more 
$z>2$ quasar candidates than SDSS even at the same spectroscopic magnitude 
limit. We believe
that at least some SDSS missing quasars with $z\sim2.7$ can be recovered with 
our selection criteria. Deeper spectroscopic observations on the quasar
candidates selected with our criteria are expected to confirm our results 
and produce a more complete and much larger quasar sample in the future.

The quasar selection criteria we proposed in this paper are fairly simple, involving
only cuts in the two-colour diagrams for optical point sources. More complicated
approaches, such as the multi-dimensional search, would be more optimal. To
check this, we have
tried several different ways to search the criteria for separating quasars and stars,
including 3-dimensional colour-colour diagrams, support vector machine (SVM) and
Bayesian classification algorithm, but we did not find significant improvements. 
The detailed comparisons of the results by adopting these different 
approaches with our current work will be presented in a separated paper. 

The study presented in this paper is mainly based on the SDSS-UKIDSS sample of
8498 quasars. Although the cross-identifications were done by finding the closest
UKIDSS counterparts of the SDSS quasars and the offsets between the SDSS and UKIDSS
positions are mostly within 0.5$''$, 
the mis-identifications of some sources are unavoidable. However, by checking the 
colour-z relations
and colour-colour diagrams of these SDSS-UKIDSS quasars, we believe that the fraction of
such mis-identifications should be very low. Therefore, the main results obtained 
in this paper will not be affected by the mis-identifications of some sources. 

Because there is no quasar with $z>5.3$ in our SDSS-UKIDSS quasar sample, our
quasar selection criteria can only be applied to identify $z<5.3$ quasars. 
Although we present a criterion in the $Y-J$ vs $i-z$ diagram for finding 
$z>5.6$ quasars based on the colour-z relation given by Hewett et al. (2006), 
this criterion obviously needs to be tested with a large sample
of $z>5.6$ quasars with both SDSS and UKIDSS data. Moreover, for the high-z
quasars with $z>5$, the photometric redshift estimation will largely rely on
the UKIDSS colours because both $u-g$ and $g-r$ colours are not available. We 
also need a sample of $z>5$ quasars with both SDSS and UKIDSS photometric data
to check the accuracy of our photometric redshift estimations. This is expected
to be done in the future when a larger high-z quasar sample is available. 

Although we made two tests using a UKIDSS/EDR sample and a sample of variable sources
in SDSS Stripe 82 to check the robustness of our quasar selection criteria and the
photometric redshift method, obviously we still need more tests. We have selected some
quasar candidates with $i_{\rm{AB}}<19.1$ using our criteria from the variable sources 
in Stripe 82, which do not have SDSS spectroscopy, and will identify them with optical
spectroscopy. This will provide direct evidence for whether using our SDSS/UKIDSS selection
criteria can help to find missing quasars in SDSS. On the other hand, finding quasars with 
certain criteria and comparing them with those in the existing spectroscopic surveys may not 
be enough for testing the completeness, as most known quasars were identified with the colour 
selection methods.  A better way to do it may be to run  
simulated colours of quasars, with assumptions on their continuum,  
emission line and reddening, for different redshift and luminosity (Fan 1999),  
and check how many simulated quasars can be found with the selection criteria. This
will be studied in our future work.     

The Chinese LAMOST quasar survey will be done in the next a few years with the
most efficient spectroscopic telescope in the world, and is expected to obtain
a much larger quasar sample (with the magnitude limit of $i$=20.5 or 21) than the 
previous surveys. The results of our current study, if confirmed with further spectroscopic
observations, will be helpful for preparing the 
input catalog of quasar candidates and estimating their photometric redshifts for
the LAMOST quasar survey, especially in the overlapped sky area between SDSS and 
UKIDSS. By doing necessary tests of various selection criteria with the spectroscopic
observations in the pilot survey, we will determine the best criteria for quasar 
candidates selections and apply them to the LAMOST quasar survey. Although the 
UKIDSS/LAS will cover only a sky area of 4000deg$^2$, with the near-IR colours we will
be able to select more complete quasar samples with redshift up to 5 than previous 
surveys. This large and
complete quasar sample will be very important to many further studies such as
those on the quasar luminosity function, quasar clustering, and large scale structure 
in the universe.

\section*{Acknowledgments}
We thank Tinggui Wang, Weimin Yuan, Yanxia Zhang and Hongyan Zhou for 
stimulating discussions, and Xiaohui Fan for helpful suggestions 
and comments.
We are grateful to the referee, Stephen Warren, for a very constructive
report, which helps us to improve the paper substantially.
The work is supported by the National 
Natural Science Foundation of China  (10525313) and the
National Key Basic Research Science Foundation of China (2007CB815405).

Funding for the SDSS and SDSS-II has been provided by the Alfred P. Sloan Foundation, 
the Participating Institutions, the National Science Foundation, the U.S. Department of 
Energy, the National Aeronautics and Space Administration, the Japanese Monbukagakusho, 
the Max Planck Society, and the Higher Education Funding Council for England. The SDSS 
Web Site is http://www.sdss.org/.

The SDSS is managed by the Astrophysical Research Consortium for the Participating 
Institutions. The Participating Institutions are the American Museum of Natural History, 
Astrophysical Institute Potsdam, University of Basel, University of Cambridge, Case 
Western Reserve University, University of Chicago, Drexel University, Fermilab, the 
Institute for Advanced Study, the Japan Participation Group, Johns Hopkins University, 
the Joint Institute for Nuclear Astrophysics, the Kavli Institute for Particle 
Astrophysics and Cosmology, the Korean Scientist Group, the Chinese Academy of Sciences 
(LAMOST), Los Alamos National Laboratory, the Max-Planck-Institute for Astronomy (MPIA), 
the Max-Planck-Institute for Astrophysics (MPA), New Mexico State University, Ohio State 
University, University of Pittsburgh, University of Portsmouth, Princeton University, 
the United States Naval Observatory, and the University of Washington.


\begin{table}
\caption{Median colour-z relations of SDSS-UKIDSS quasars}
\begin{scriptsize}
\setlength{\tabcolsep}{3.5pt}

\begin{tabular}{ccccccccc}\\ \hline
Redshift & $u-g$ & $g-r$ & $r-i$ & $i-z$ & $z-Y$ & $Y-J$ & $J-H$ & $H-K$\\
\hline

0.025&0.09&0.88&0.58&0.42&-1.87&0.54&0.76&0.61\\ 
0.075&-0.02&0.87&0.65&0.39&-1.02&0.57&0.74&0.65\\ 
0.125&-0.37&0.87&0.66&0.35&-0.69&0.58&0.74&0.73\\ 
0.175&-0.64&0.84&0.64&0.3&-0.48&0.60&0.72&0.76\\ 
0.225&-0.74&0.83&0.62&0.31&-0.30&0.59&0.76&0.85\\ 
0.275&-0.86&0.72&0.42&0.54&-0.38&0.54&0.80&0.90\\ 
0.325&-0.91&0.55&0.30&0.65&-0.40&0.51&0.80&0.95\\ 
0.375&-0.84&0.42&0.32&0.57&-0.28&0.51&0.80&0.92\\ 
0.425&-0.78&0.31&0.42&0.44&-0.24&0.51&0.77&0.95\\ 
0.475&-0.75&0.25&0.43&0.30&0.04&0.35&0.76&0.92\\ 
0.525&-0.72&0.21&0.41&0.21&0.34&0.06&0.74&0.91\\ 
0.575&-0.69&0.23&0.40&0.12&0.49&-0.02&0.75&0.88\\ 
0.625&-0.67&0.20&0.37&0.08&0.39&0.06&0.71&0.89\\ 
0.675&-0.70&0.23&0.29&0.17&-0.04&0.38&0.67&0.93\\ 
0.725&-0.71&0.26&0.20&0.27&-0.15&0.45&0.65&0.91\\ 
0.775&-0.72&0.29&0.18&0.29&-0.18&0.69&0.44&0.90\\ 
0.825&-0.77&0.30&0.16&0.31&-0.23&0.85&0.29&0.87\\ 
0.875&-0.77&0.35&0.15&0.28&-0.18&0.86&0.21&0.83\\ 
0.925&-0.84&0.38&0.14&0.25&-0.13&0.81&0.17&0.84\\ 
0.975&-0.87&0.44&0.18&0.24&-0.04&0.72&0.20&0.83\\ 
1.025&-0.88&0.48&0.16&0.20&0.03&0.40&0.42&0.83\\ 
1.075&-0.89&0.49&0.19&0.15&0.08&0.18&0.60&0.79\\ 
1.125&-0.93&0.49&0.20&0.09&0.10&0.10&0.63&0.80\\ 
1.175&-0.96&0.49&0.22&0.08&0.10&0.12&0.62&0.75\\ 
1.225&-0.96&0.50&0.23&0.10&-0.01&0.21&0.67&0.62\\ 
1.275&-0.97&0.51&0.23&0.10&0.03&0.21&0.79&0.47\\ 
1.325&-0.98&0.50&0.24&0.11&-0.02&0.27&0.81&0.39\\ 
1.375&-0.96&0.48&0.26&0.13&-0.06&0.34&0.76&0.37\\ 
1.425&-0.93&0.45&0.31&0.13&-0.10&0.42&0.71&0.32\\ 
1.475&-0.89&0.39&0.37&0.13&-0.14&0.45&0.69&0.36\\ 
1.525&-0.81&0.37&0.42&0.12&-0.14&0.46&0.68&0.37\\ 
1.575&-0.80&0.33&0.43&0.14&-0.16&0.46&0.64&0.32\\ 
1.625&-0.84&0.31&0.44&0.15&-0.15&0.44&0.64&0.34\\ 
1.675&-0.86&0.29&0.46&0.17&-0.12&0.40&0.58&0.39\\ 
1.725&-0.90&0.27&0.46&0.17&-0.12&0.35&0.51&0.50\\ 
1.775&-0.92&0.27&0.47&0.16&-0.07&0.31&0.40&0.62\\ 
1.825&-0.97&0.27&0.48&0.18&-0.05&0.28&0.39&0.63\\ 
1.875&-0.97&0.27&0.47&0.20&-0.05&0.27&0.39&0.60\\ 
1.925&-0.98&0.29&0.42&0.25&-0.10&0.25&0.40&0.62\\ 
1.975&-0.95&0.31&0.39&0.31&-0.07&0.25&0.42&0.58\\ 
2.025&-0.94&0.32&0.35&0.34&-0.06&0.22&0.47&0.56\\ 
2.075&-0.90&0.34&0.34&0.36&-0.10&0.20&0.51&0.63\\ 
2.125&-0.80&0.34&0.32&0.38&-0.06&0.21&0.53&0.74\\ 
2.175&-0.70&0.35&0.32&0.38&-0.03&0.25&0.52&0.83\\ 
2.225&-0.60&0.34&0.31&0.38&-0.06&0.25&0.53&0.85\\ 
2.275&-0.50&0.30&0.28&0.37&-0.07&0.28&0.53&0.80\\ 
2.325&-0.44&0.25&0.30&0.37&-0.02&0.32&0.50&0.84\\ 
2.375&-0.43&0.28&0.25&0.39&-0.01&0.34&0.56&0.80\\ 
2.425&-0.45&0.34&0.26&0.41&0.04&0.29&0.54&0.77\\ 
2.475&-0.43&0.31&0.24&0.43&0.06&0.28&0.54&0.80\\ 
2.525&-0.45&0.34&0.23&0.38&0.06&0.27&0.55&0.78\\ 
2.575&-0.45&0.33&0.23&0.32&0.09&0.24&0.41&0.76\\ 
2.625&-0.44&0.35&0.24&0.39&0.09&0.26&0.46&0.57\\ 
2.675&-0.21&0.35&0.27&0.33&0.15&0.28&0.45&0.53\\ 
2.725&-0.14&0.40&0.35&0.39&0.13&0.29&0.50&0.58\\ 
2.775&0.10&0.50&0.24&0.31&0.19&0.32&0.49&0.46\\ 
2.825&0.14&0.39&0.30&0.24&0.23&0.34&0.43&0.48\\ 
2.875&0.33&0.40&0.28&0.25&0.16&0.37&0.41&0.49\\ 
2.925&0.37&0.46&0.30&0.30&0.19&0.40&0.43&0.53\\ 
2.975&0.60&0.47&0.29&0.23&0.11&0.43&0.45&0.50\\ 
3.05&0.61&0.48&0.31&0.22&0.10&0.41&0.45&0.46\\ 
3.15&1.23&0.50&0.30&0.20&0.13&0.46&0.44&0.56\\ 
3.25&2.29&0.57&0.28&0.20&0.14&0.50&0.43&0.64\\ 
3.35&4.24&0.70&0.32&0.23&0.06&0.55&0.46&0.68\\ 
3.45&6.67&0.94&0.35&0.25&-0.06&0.59&0.41&0.66\\ 
3.55&---&1.13&0.35&0.21&-0.03&0.57&0.44&0.60\\ 
3.65&---&1.34&0.31&0.25&-0.05&0.63&0.52&0.63\\ 
3.75&---&1.41&0.29&0.25&-0.01&0.59&0.56&0.59\\ 
3.85&---&1.55&0.31&0.29&0.01&0.55&0.60&0.67\\ 
3.95&---&1.49&0.37&0.23&-0.08&0.58&0.59&0.58\\ 
4.05&---&1.76&0.38&0.24&-0.05&0.46&0.62&0.53\\ 
4.15&---&1.92&0.49&0.23&0.03&0.43&0.64&0.49\\ 
4.25&---&2.10&0.66&0.27&-0.07&0.42&0.61&0.59\\ 
4.35&---&2.00&0.63&0.32&0.05&0.26&0.65&0.46\\ 
4.45&---&2.25&0.59&0.31&0.10&0.38&0.84&0.48\\ 
4.60&---&2.38&1.32&0.30&0.05&0.30&0.75&0.48\\ 
4.80&---&3.87&1.57&0.19&0.05&0.33&0.78&0.71\\ 
5.00&---&7.37&1.81&0.33&-0.19&0.38&0.68&0.54\\ \hline
\end{tabular}
\end{scriptsize}
\end{table}
\begin{table}
\caption{Table 2 continued}
\label{Table 2 continued}
\begin{scriptsize}
\setlength{\tabcolsep}{3.5pt}
\begin{tabular}{ccccccccc}\\ \hline
Redshift & $u-g$ & $g-r$ & $r-i$ & $i-z$ & $z-Y$ & $Y-J$ & $J-H$ & $H-K$\\
\hline

5.20&---&---&1.44&0.73&0.16&0.40&0.612&0.54\\ 
5.30&---&---&1.58&0.82&0.23&0.36&0.57&0.58\\ 
5.40&---&---&1.73&0.96&0.33&0.29&0.53&0.63\\ 
5.50&---&---&1.95&1.14&0.37&0.26&0.51&0.67\\ 
5.60&---&---&2.13&1.48&0.34&0.27&0.49&0.69\\ 
5.70&---&---&2.05&2.07&0.29&0.29&0.47&0.70\\ 
5.80&---&---&2.06&2.62&0.24&0.32&0.46&0.72\\ 
5.90&---&---&2.47&2.86&0.24&0.33&0.46&0.73\\ 
6.00&---&---&2.99&3.04&0.26&0.35&0.50&0.73\\ 
6.10&---&---&3.57&3.18&0.33&0.39&0.52&0.74\\ 
6.20&---&---&4.34&3.32&0.47&0.42&0.52&0.76\\ 
6.30&---&---&5.56&3.46&0.67&0.43&0.50&0.79\\ 
6.40&---&---&7.61&3.62&0.88&0.44&0.48&0.81\\ \hline
\end{tabular}
Note: All colours are derived from magnitudes in Vega system. The $u-g$ colour at 
$3<z<3.5$, $g-r$ colour at $4.3<z<5.0$ and other colours at $5<z<6.4$ are adopted
from Table 26 of Hewett et al. (2006).
\end{scriptsize}
\end{table}


\begin{thebibliography}{99}
\bibitem[]{}Abazajian K. et al., 2009, ApJS, 182, 543
\bibitem[]{}Ball N. M. et al., 2007, ApJ, 663, 774
\bibitem[]{}Boyle B.J. et al., 2000, MNRAS, 317, 1014
\bibitem[]{}Casali M. et al., 2007, A\&A, 467, 777
\bibitem[]{}Chiu K., Richards G.T., Hewett P.C., Maddox N., 2007, MNRAS, 375,1180
\bibitem[]{}Croom  S.M. et al., 2004, MNRAS, 349, 1397
\bibitem[]{}Croom S.M. et al., 2009, MNRAS, 392, 19
\bibitem[]{}Fan X. 1999, AJ, 117, 2528
\bibitem[]{}Fan X. et al., 2001a, AJ, 121, 54
\bibitem[]{}Fan X. et al., 2001b, AJ, 122, 2833
\bibitem[]{}Hambly N. et al., 2008, MNRAS, 384, 637
\bibitem[]{}Hewett P.C., Warren S.J., Leggett S.K., Hodgkin S.T. 2006, MNRAS, 367, 454
\bibitem[]{}Ivezic Z., Vivas A.K., Lupton R.H., Zinn R. 2005, AJ, 129, 1096
\bibitem[]{}Ivezic Z. et al., 2007, AJ, 134, 973
\bibitem[]{}Kaiser N. et al., 2007, Proc. SPIE, 4836, 154
\bibitem[]{}Lawrence A. et al., 2007, MNRAS, 379, 1599
\bibitem[]{}Maddox N., Hewett P.C., Warren S.J., Croom S.M. 2008, MNRAS, 386, 1605
\bibitem[]{}Maddox N., Hewett P.C. 2006, MNRAS, 367, 717
\bibitem[]{}Richards G.T. et al., 2001, AJ, 122, 1151
\bibitem[]{}Richards G.T. et al., 2002, AJ, 123, 2945
\bibitem[]{}Richards G.T. et al., 2006, AJ, 131, 2766
\bibitem[]{}Richards G.T. et al., 2009a, ApJS, 180, 67
\bibitem[]{}Richards G.T. et al., 2009b, AJ, 137, 3884
\bibitem[]{}Schlegel D.J., Finkbeiner D.P., Davis M. 1998, ApJ, 500. 525
\bibitem[]{}Schneider D.P. et al., 2007, AJ, 134, 102
\bibitem[]{}Sesar B. et al., 2007, AJ, 134, 2236 
\bibitem[]{}Smith J.R. et al.,  2005, MNRAS, 359, 57
\bibitem[]{}Su D.Q., Cui X., Wang Y., Yao Z.  1998, Proc. SPIE, 3352, 76
\bibitem[]{}Tyson J.A. 2002, Proc. SPIE, 4836, 10
\bibitem[]{}Warren S.J., Hewett P.C., Foltz C.B. 2000, MNRAS, 312, 827
\bibitem[]{}Weinstein M.A. et al., 2004, ApJS, 155, 243
\bibitem[]{}Wu X.-B., Zhang W., Zhou X. 2004,ChJAA, 4, 17
\bibitem[]{}York D.G. et al., 2000, AJ, 120,1579

\end{thebibliography}
\end{document}